\documentclass[oneside,a4paper,12pt]{article}
\usepackage{latexsym}
\usepackage{euscript}
\usepackage{epsfig}
 \large\normalsize

\def\Frac#1#2{\frac{\displaystyle{#1}}{\displaystyle{#2}}}
\def\lsim{\raise0.3ex\hbox{$\;<$\kern-0.75em\raise-1.1ex\hbox{$\sim\;$}}}
\def\gsim{\raise0.3ex\hbox{$\;>$\kern-0.75em\raise-1.1ex\hbox{$\sim\;$}}}

\def\npb#1#2#3{    {\it Nucl. Phys. }{\bf B #1} (#2) #3}

\def\plb#1#2#3{    {\it Phys. Lett. }{\bf B #1} (#2) #3}
\def\prd#1#2#3{    {\it Phys. Rev. }{\bf D #1} (#2) #3}

\def\prl#1#2#3{    {\it Phys. Rev. Lett. }{\bf #1} (#2) #3}
\def\ptp#1#2#3{    {\it Prog. Theor. Phys. }{\bf #1} (#2) #3}
    
\def\zpc#1#2#3{    {\it Zeit. f\"ur Physik }{\bf C #1} (#2) #3}

\begin{document}
\begin{flushright}
 FTUAM-00-08\\ 
 KUNS-1648\\
 SISSA/23/2000/EP
\end{flushright}
\vskip 1cm
\begin{center}
{\Large \bf EDM--free supersymmetric CP violation with non--universal 
soft terms}
\vskip 0.75cm
{Shaaban Khalil$^{1,2}$, Tatsuo Kobayashi$^3$ and Oscar
Vives$^4$}\\ \vspace*{1cm} \small{\textit{$^1$Departmento de   
Fisica Te\'orica, C.XI, Universidad Aut\'onoma de Madrid,\\ 28049
Cantoblanco, Madrid, Spain.}} \\
\vspace*{2mm}\small{\textit{$^2$Ain Shams University, Faculty of
Science, Cairo 11566, Egypt.}} \\
\vspace*{2mm}\small{\textit{$^3$ Department of Physics, Kyoto University,
Kyoto 606-8502, Japan}}\\
\vspace*{2mm}\small{\textit{$^4$SISSA -- ISAS, via Beirut 2-4, 34013,
Trieste, Italy and} \\
\textit{INFN, Sezione di Trieste, Trieste, Italy}}\\
\vspace*{1.5cm}

\begin{abstract}
{Non--universality in the soft breaking terms is a common feature in most 
superstring inspired SUSY models. This property is required to obtain 
sizeable CP violation effects from SUSY and, on the 
other hand, can be used to avoid the Electric Dipole Moment constraints. 
We take advantage of these qualities and explore a class 
of SUSY models based on type I string theory where scalar masses, gaugino 
masses and trilinear couplings are non--universal. In this framework, we show 
that, in the presence of large SUSY phases, the bounds on the Electric Dipole 
Moments can be controled without fine--tuning. At the same time, we find that 
these phases, free from EDM constraints, lead to large contributions to the 
observed CP phenomena in Kaon system and, in particular, to direct CP 
violation in $\varepsilon'/\varepsilon$.}
\end{abstract}

\setcounter{page}{1}
\end{center}

\thispagestyle{empty}

\newpage
\section{Introduction}

CP violation constitutes one of the main open questions in high energy 
physics at the beginning of the 21st century. The Standard Model (SM) of 
electroweak interactions is able to accommodate the experimentally 
observed CP violation through a phase, $\delta_{CKM}$, in the 
Cabibbo--Kobayashi--Maskawa (CKM) mixing matrix. In spite of this, there 
exist strong hints from other fields (for instance, electroweak baryogenesis)
that suggest that this can not be the only source of CP violation 
present in nature.
\vskip 0.25cm

In fact, most of the extensions of the SM include new phases that may modify 
the SM predictions in CP violation phenomena. For example, even in the 
simplest supersymmetric extension of the SM, the so--called Constrained
Minimal Supersymmetric Standard Model (CMSSM), we have new phases in the 
gaugino masses, $A$--terms and the $\mu$--term \cite{2phases}. However, it 
is known since the early 80s \cite{edm-susy}, that the presence of these 
phases for SUSY masses around the electroweak scale gives rise to 
supersymmetric contributions to the Electric Dipole Moment (EDM) of the 
electron and the neutron roughly two 
orders of magnitude above the experimental bounds. Hence, given these
strong constraints, most of the people working in SUSY phenomenology take 
these phases as exactly vanishing. Although this might be the most direct
choice, it has been recently shown that there exist some other mechanisms which
allow large SUSY phases while respecting EDM bounds.
For instance, one of these mechanisms consists in a possible destructive 
interference among different contributions to the EDM that can occur in 
some regions of the SUSY parameter space~\cite{cancel,pokorski,brhlik2}. 
A second option is to have approximately degenerate heavy 
sfermions for the first two generations~\cite{effective} and finally, a 
third possibility (and maybe more natural) is to have non--universal
soft supersymmetry breaking terms~\cite{abel,masiero,khalil,non-u}.
\vskip 0.25cm

In the presence of one of these mechanisms, large CP phases will be present
in the SUSY sector and one may expect important effects on CP physics 
other than the EDM, e.g. in the $K$ and $B$ systems. 
However, in Ref.~\cite{demir}, it has been shown that, for vanishing 
$\delta_{CKM}$, a general SUSY model with all possible phases in 
the soft--breaking terms, but no new flavor structure beyond the usual 
Yukawa matrices can never generate a sizeable contribution to $\varepsilon_K$, 
$\varepsilon'/\varepsilon$ or hadronic $B^0$ CP asymmetries.
This means that the presence of non--universal soft breaking terms besides 
large SUSY phases is crucial to enhance these CP violation effects.
In agreement with this, it has been explicitly shown 
that contributions to $\varepsilon_K$ are small within the 
dilaton--dominated SUSY breaking of the weakly coupled heterotic string
model~\cite{barr}, where $A$--terms as well as gaugino masses are universal.
On the other hand, it is well--known that the strict universality in the soft
breaking sector is a strong assumption not realized in many supergravity and 
string inspired models~\cite{strings}.
All these arguments indicate not only that the presence of non--universal 
soft terms can solve the problem of too large contributions to EDMs but also
that it allows for large SUSY contributions in CP violation experiments. 
Hence, in this work we will follow this avenue and analyze the effects of 
non--universal soft terms in both EDM and CP violation in the $K$--system.   
\vskip 0.25cm

In particular, non--universality of $A$--terms has been shown to be very
effective to generate large CP violation effects 
\cite{abel,masiero,khalil,non-u,kane}. In fact, the presence of 
non--degenerate $A$--terms is essential for enhancing the gluino contributions 
to $\varepsilon'/\varepsilon$ through large imaginary parts of the $L$--$R$ 
mass insertions, $\mathrm{Im}(\delta_{LR})_{12}$ and 
$\mathrm{Im}(\delta_{RL})_{12}$, as recently emphasized in Ref. \cite{masiero}.
These SUSY contributions can, indeed, account for a sizeable part of the 
recently measured experimental value of $\varepsilon'/\varepsilon$ 
\cite{KTeV,NA31}. In the following, we will present an explicit realization
of such mechanism in the framework of a type I superstring inspired SUSY 
model. Within this model, it is possible to obtain non--universal soft
breaking terms, i.e. scalar masses, gaugino masses and trilinear
couplings. We show that here EDMs can be sufficiently small
while the SUSY phases of the off diagonal $A$--terms are large, and enough
to generate sizable contribution to $\varepsilon'/\varepsilon$. 
\vskip 0.25cm

This paper is organized as follows. In section 2 we show our starting
models based on type I string theory. We emphasize that it is quite natural
to obtain non--universal $A$--terms in these models. In section 3 we discuss
the impact of these new flavor structures on the sfermion mass matrices. We 
show that non--universality of the $A$--terms, in particular, can generate 
sizable off--diagonal entries in the down squark mass matrix. Section 4 is 
devoted to the discussion of the constraints from the EDMs of the electron 
and neutron. We explain that, in the model we consider with non--universal 
$A$--terms, the EDMs can be kept sufficiently small while there are still two 
phases completely unconstrained. The effect of these two phases in other CP
violating process are given in section 5, where we study explicitly the
effect of these phases in the K--system. Finally, in section 6, we give our 
conclusions.

\section{Type I models}

In this section we explain our starting model, which is
based on type I string models.
The purpose of the paper is to study explicitly several CP aspects
in
models with non--universal soft breaking terms.
Type I models can realize such initial conditions, in particular,
it is
possible to obtain non--universality in the scalar masses, 
$A$--terms
and gaugino masses\footnote{Different possibilities can be found
in \cite{ibanez}.}. To obtain non--universal gaugino masses,
we must assign the gauge groups to different branes \cite{brhlik2,
ibrahim}.
Type I models contain nine--branes and three types of 
five--branes ($5_a$, $a=1,2,3$).
Phenomenologically there is no difference between 
the 9--brane and the 5--branes.
A gauge multiplet is assigned on one set of branes.
Only if the SM gauge group is not associated with a single set
of branes, the gaugino masses can be non--universal. 
Here we assume that the gauge group $SU(3)$ on one of the branes and 
the gauge group $SU(2)$ on another brane. 
We call these branes the $SU(3)$--brane and the $SU(2)$--brane, 
respectively.
\vskip 0.25cm

Now we assign chiral matter fields and the brane corresponding 
to $U(1)_Y$ such that we obtain non--universal $A$--terms.
Chiral matter fields correspond to open strings spanning between 
branes. Thus, chiral matter fields have non--vanishing 
quantum numbers only for the gauge groups corresponding to 
the branes between which the open string spans.
For example, the chiral field corresponding to the 
open string between the $SU(3)$ and $SU(2)$ branes can have 
non--trivial representations under both $SU(3)$ and $SU(2)$, 
while the chiral field corresponding to the open string, 
which starts and ends on the $SU(3)$--brane, should be 
an $SU(2)$--singlet. Furthermore, it is required that $U(1)_Y$ should
correspond to one of the $SU(3)$--brane and $SU(2)$--brane but not another
brane such that quark doublets have non--vanishing $U(1)_Y$ charges.
\vskip 0.25cm

While there is only one type 
of the open string which spans between different branes, 
there are three types of open strings which start and end on the 
same brane, that is, the $C_i$ sectors (i=1,2,3),
which corresponding to the $i$-th complex compact dimension 
among the three complex dimensions.
If we assign the three families to the different $C_i$ sectors 
each other, we obtain non--universality.
That is the only possible non--universality and 
it is important for model building.
That implies that we can not derive 
non--universality for the squark doublets, i.e. 
the left--handed sector.
Non--universality can appear in the right--handed sector 
only if $U(1)_Y$ corresponds to the $SU(3)$--brane
and the families are assigned to different $C_i$ sectors each
other.\footnote{It is possible to assign $U(1)_Y$ as 
a linear combination of $U(1)$ symmetry on the $SU(3)$--brane and 
$U(1)$ symmetries on other branes including the $SU(2)$--brane.
However, in this case, phenomenological results are same.}
Therefore, the model leading to both 
non--universal gaugino masses and non--universal $A$--terms is 
unique, that is, we assign $SU(3)\times U(1)_Y$ and $SU(2)$ to different
branes. The quark doublets correspond to the open string between 
the $SU(3)\times U(1)_Y$--brane and the $SU(2)$--brane.
The quark singlets correspond to three different sectors on the 
$SU(3)$--brane.
Hence, non--universality of soft SUSY terms can appear 
only for the right--handed sector, while soft SUSY breaking terms 
are universal for the left--handed sector.
\vskip 0.25cm

Here we assume that the gauge
group $SU(3)\times U(1)_Y$ is originated from the 9--brane and the
gauge group $SU(2)$ is originated from the $5_1$--brane 
like Ref.~\cite{brhlik2,ibrahim}.\footnote{
Different assignment of branes lead to phenomenologically 
similar results as emphasized above.}
In this case 
$SU(2)$--doublet fields, e.g. quark doublets and the Higgs fields,
should be assigned to the open string, which spans between the
$5_1$ and 9--branes and is denoted by the $C^{95_1}$ sector. 
On the other hand, the $SU(2)$--singlet fields, e.g. 
quark singlets, correspond to the open string, which starts and ends 
on the 9--brane.
Such open string includes three sectors  denoted by $C^{9}_i$  
($i=1,2,3$).
\vskip 0.25cm

At the string level, only the $C^{9}_1$ sector is allowed 
in the 3--point $C^{95_1}C^{95_1}C^{9}_1$ coupling.
However, we assume that the Yukawa couplings for 
the other sectors $C^{9}_i$ ($i=1,2$) are allowed 
through higher dimensional operators after symmetry breaking 
within the framework of effective field theory.
Such effective Yukawa couplings originated 
from higher dimensional operators naturally lead to 
suppressed values of couplings, 
while the $C^{95_1}C^{95_1}C^{9}_1$ coupling 
would correspond to the large Yukawa coupling of the top quark 
as well as the bottom quark.
Within such framework, the hierarchical structure of fermion mass
matrices could be realized.
Then we allow all of the $C^{9}_i$  ($i=1,2,3$) 
as candidates of quark singlets.
In particular, we assign the $C^{9}_1$ sector 
to the third family.
Also we assign the first and second families of 
quark singlets to $C^{9}_3$ and $C^{9}_2$, respectively,   
in order to derive non--universal A--terms.
\vskip 0.25cm

Under the above assignment of the gauge multiplets and 
the matter fields, soft SUSY breaking terms are obtained,  
following the formulae in Ref.~\cite{ibanez}.
The gaugino masses are obtained 
\begin{eqnarray}
\label{gaugino}
M_3 & = & M_1 = \sqrt 3 m_{3/2} \sin \theta\  e^{-i\alpha_S}, \\
M_2 & = &  \sqrt 3 m_{3/2} \cos \theta\ \Theta_1 e^{-i\alpha_1}.
\end{eqnarray}
While the $A$-terms are obtained as 
\begin{equation}
A_{C_1^9}= -\sqrt 3 m_{3/2} \sin \theta\ e^{-i\alpha_S}=-M_3,
\label{A-C1}
\end{equation}
for the coupling including $C_1^{9}$, i.e. the third family, 
\begin{equation}
A_{C_2^9}= -\sqrt 3 m_{3/2}(\sin \theta\ e^{-i\alpha_S}+
\cos \theta\ (\Theta_1 e^{-i\alpha_1}- \Theta_2 e^{-i\alpha_2})),
\label{A-C2}
\end{equation}
for the coupling including $C_2^{9}$, i.e. the second 
family  and 
\begin{equation}
\label{A-C3}
A_{C_3^9}= -\sqrt 3 m_{3/2}(\sin \theta\ e^{-i\alpha_S}+
\cos \theta\ (\Theta_1 e^{-i\alpha_1}- \Theta_3 e^{-i\alpha_3})),
\end{equation}
for the coupling including $C_3^{9}$, i.e. the first family.
Here $m_{3/2}$ is the gravitino mass, $\alpha_S$ and $\alpha_i$ are 
the CP phases of the F-terms of the dilaton field $S$ and 
the three moduli fields $T_i$, and $\theta$ and $\Theta_i$ are 
goldstino angles, and we have the constraint, $\sum \Theta_i^2=1$.
\vskip 0.25cm 

Thus, if quark fields correspond to different 
$C_i^9$ sectors, we have non--universal A--terms.
Then we obtain the following A--matrix for both of the 
up and down sectors, 
\begin{eqnarray}
A= \left(
\begin{array}{ccc}
A_{C^9_3}  & A_{C^9_2} & A_{C^9_1} \\ A_{C^9_3} & A_{C^9_2} &
A_{C^9_1} \\ A_{C^9_3} & A_{C^9_2} & A_{C^9_1}
\end{array}
\right) \label{A-1}.
\end{eqnarray}
Note that the non--universality appears 
only for the right--handed sector.
The trilinear SUSY breaking matrix, $(Y^A)_{ij}=(Y)_{ij}(A)_{ij}$, 
itself is obtained 
\begin{equation}
\label{trilinear}
Y^A = \left(\begin{array}{ccc}
 &  &  \\  & Y_{ij} &  \\  &  & \end{array}
\right) \cdot 
\left(\begin{array}{ccc}
A_{C^9_3} & 0 & 0 \\ 0 & A_{C^9_2} & 0 \\ 0 & 0 & A_{C^9_1} \end{array}
\right),
\end{equation}
in matrix notation.
\vskip 0.25cm

In addition, soft scalar masses for quark doublets and 
the Higgs fields are obtained, 
\begin{equation}
\label{doublets}
m_{C^{95_1}}^2=m_{3/2}^2(1-\Frac{3}{2} \cos^2 \theta\ (1- 
\Theta_1^2)).
\end{equation}
The soft scalar masses for quark singlets are obtained as
\begin{equation}
\label{singlets}
m_{C_i^9}^2=m_{3/2}^2(1-3\cos^2 \theta\ \Theta^2_i),
\end{equation}
if it corresponds to  the $C_i^{9}$ sector.
\vskip 0.25cm

Finally, we fix the magnitudes of the $\mu$--term and $B$--term by using 
the minimization conditions of the Higgs potential. This completes the
whole set of initial conditions in our type I string inspired model.

\section{Flavor physics and soft breaking terms}
\label{sec:flavor}

In the previous section, we have defined our string inspired model. Below
the string scale, this model is simply a MSSM
(understood as with the minimal particle content from the SM) with 
non--trivial soft--breaking terms from the point of view of flavor. 
Scalar mass matrices and trilinear terms have completely new flavor 
structures, as opposed to the supergravity inspired CMSSM or the SM, 
where the only connection 
between different generations is provided by the Yukawa matrices.
\vskip 0.25cm

This model includes in the quark sector 7 different structures of flavor,
$M_{Q}^2$, $M_{U}^2$, $M_{D}^2$, $Y_d$, $Y_u$, $Y^A_d$ and $Y^A_u$ 
($(Y^A_q)_{i j} = (A_q)_{i j} (Y_q)_{i j}$). From these matrices, $M_{Q}^2$,
the squark doublet mass matrix, is proportional to the identity matrix, and 
hence trivial, then we are left with 6 non--trivial flavor matrices.
Notice that we have always the freedom to diagonalize the hermitian
squark mass matrices and then Yukawa and trilinear matrices 
are completely fixed. This implies that, in this case, these four matrices
are observable, as opposed to the SM or CMSSM case, where only
quark masses and the CKM matrix are observable.
\vskip 0.25cm

At this point, to specify completely the model, we need not only the 
soft--breaking terms but also the complete Yukawa textures.
The only available experimental information is the CKM mixing matrix 
and the quark masses. In this work, as an estimate of 
possible effects we choose our Yukawa texture following two simple 
assumptions : i) the CKM mixing matrix originates from the down Yukawa 
couplings and ii) Yukawa matrices are hermitian \cite{RRR}.
With these two assumptions we fix completely the Yukawa matrices at the 
string scale, $M_X$,  
\begin{eqnarray}
\begin{array}{lr}
Y_u=\ \Frac{1}{v_2} \left(\begin{array}{ccc}
m_u & 0 & 0 \\ 0 & m_c & 0 \\ 0 & 0 & m_t \end{array}
\right)\ \  & \ \ \ \ 
Y_d=\ \Frac{1}{v_1}\  K^\dagger\ \cdot \left(\begin{array}{ccc}
m_d & 0 & 0 \\ 0 & m_s & 0 \\ 0 & 0 & m_b \end{array}
\right) \cdot \ K
\label{Yuk}
\end{array}
\end{eqnarray}
with $v= v_1 /(\cos \beta) = v_2 / (\sin \beta) = \sqrt{2} M_W / g$.
Through
all the paper we fix $\tan \beta = v_2/v_1 =2$ and $K$ is the CKM matrix. 
In principle, generic Yukawa matrices in this basis of diagonal squark
masses could be different \cite{RRR}, but other matrices lead to physically 
similar results for the following analyses. Hence, the texture in 
Eq.(\ref{Yuk}) is enough for our purposes. 
\vskip 0.25cm

In this basis we can analyze the down trilinear matrix that with 
Eqs.(\ref{trilinear}) and (\ref{Yuk}) is,
\begin{equation}
Y^A_d (M_{X}) = \Frac{1}{v_1}\  K^\dagger\ \cdot \ M_d\ \cdot \ K\ \cdot
\left(\begin{array}{ccc}
A_{C^9_3} & 0 & 0 \\ 0 & A_{C^9_2} & 0 \\ 0 & 0 & A_{C^9_1} \end{array}
\right)
\end{equation}
with $M_d=diag.(m_d, m_s, m_b)$.
\vskip 0.25cm

Hence, together with the up trilinear matrix we have our model completely
defined. The next step is simply to use the MSSM Renormalization Group 
Equations (RGE) \cite{RGE} to obtain the whole spectrum and couplings at the
low scale, $M_W$. The dominant effect in RGEs of the trilinear terms 
is due to the gluino mass which produces the well--known alignment among
A--terms and gaugino phases. However this RG effect is always 
proportional to the Yukawa matrices and not to the trilinear terms themselves,
that is, roughly the RGEs are $ d Y^A_d/d t \sim F(\alpha_s)\ m_{\tilde{g}}
\cdot Y_d + G(\alpha_s,\alpha_W, Y_d, Y_u,\dots)\cdot  Y^A_d$ \cite{RGE}.
This implies that, in the SCKM basis the gluino effects 
are diagonalized in excellent approximation, while due to the different
flavor structure of the trilinear terms large off--diagonal elements 
remain with phases ${\mathcal{O}}(1)$ \cite{masiero}. To see this more 
explicitly, we can roughly approximate the RGE effects as,
\begin{equation}
\label{YArge}
Y^A_d (M_{W}) = c_{\tilde{g}}\  m_{\tilde{g}} \ Y_d \ +c_{A}\ Y_d\ \cdot
\left(\begin{array}{ccc}
A_{C^9_3} & 0 & 0 \\ 0 & A_{C^9_2} & 0 \\ 0 & 0 & A_{C^9_1} \end{array}
\right)
\end{equation}
with $m_{\tilde{g}}$ the physical gluino mass and $c_{\tilde{g}}$, 
$c_A$ coefficients order 1 (typically $c_{\tilde{g}}\simeq 5$, $c_A\simeq 1$).
\vskip 0.25cm

We go to the SCKM basis after diagonalizing all the Yukawa matrices
(i.e. $K \cdot Y_d \cdot K^\dagger = M_d /v_1$).
In this basis we obtain the trilinear couplings as,
\begin{equation}
v_1\ Y^A_d (M_{W}) = \Big(c_{\tilde{g}}\ m_{\tilde{g}}\ M_d \ +
c_{A}\ M_d\ \cdot \ K\ \cdot \left(\begin{array}{ccc}
A_{C^9_3} & 0 & 0 \\ 0 & A_{C^9_2} & 0 \\ 0 & 0 & A_{C^9_1} \end{array}
\right)\ \cdot \ K^\dagger \Big).
\label{A-SCKM}
\end{equation}  
{}From this equation we can get the $L$--$R$ down squark mass matrix
\begin{equation} 
{M_{LR}^{(d)}}^{2}=v_1\ {Y^A_{d}}^* - \mu e^{i\varphi_{\mu}}
\tan\beta\, M_{d}.
\end{equation}
Finally using unitarity of $K$ we obtain for the $L$--$R$ 
mass insertions,
\begin{eqnarray}
\label{DLR}
(\delta_{LR}^{(d)})_{i j}= \frac{1}{m^2_{\tilde{q}}}\ m_i\ \Big(
\delta_{ij}\ (c_{A} A_{C^9_3}^*\ +\ c_{\tilde{g}}\ m_{\tilde{g}}^* -\ 
\mu e^{i\varphi_{\mu}} \tan\beta ) + \nonumber \\
K_{i 2}\ K^*_{j 2}\ c_{A}\ ( A_{C^9_2}^* - A_{C^9_3}^* ) +
K_{i 3}\ K^*_{j 3}\ c_{A}\ ( A_{C^9_1}^* - A_{C^9_3}^* ) \Big)
\end{eqnarray}
where $m^2_{\tilde{q}}$ is an average squark mass and $m_i$ the quark mass.
\vskip 0.25cm

This expression shows the main effects of the non--universal $A$--terms.
In the first place, we can see that the diagonal elements are still very 
similar
to the universal $A$--terms situation. Apart of the usual scaling with the
quark mass, these flavor--diagonal mass insertions receive 
dominant contributions from the corresponding $A_{C^9_i}$ terms 
(due to the fact that the CKM mixing matrix is close to the identity)
plus an approximately equal contribution from gluino to all three generations
and an identical $\mu$ term contribution. Hence, given that the gluino 
RG effects are dominant in Eq.(\ref{YArge}), also the phases of 
these terms tend to align with the gluino phase, as in the CMSSM. Therefore, 
EDM bounds constrain mainly the relative phase between $\mu$ and gluino 
(or chargino) and give a relatively weaker constraint to the relative 
phase between $A_{C^9_3}$ (the first generation $A$--term) and the relevant
gaugino \cite{CPuniv} as we will show in the next section. Effects of 
different $A_{C^9_i}$ in these elements are suppressed by squared CKM
mixing angles. However, flavor--off--diagonal elements are completely new 
in our model. They do not receive significant contributions from gluino 
nor from $\mu$ and so their phases are still determined by the $A_{C^9_i}$
phases and, in principle, they do not directly contribute to EDMs . 
It is also important to notice that in these off--diagonal elements the 
relevant quark mass is the one 
of the left--handed quark, see Eqs.(\ref{A-SCKM},\ref{DLR}). In section 
\ref{sec:kaon}, we will analyze the effects of these mass insertions
in the kaon system.
\vskip 0.25cm

In the same way, we must also apply the same rotations to the $L$--$L$ and 
$R$--$R$ squark mass matrices,
\begin{eqnarray}
\label{hermitian-matrices}
{M^{(d)}_{LL}}^2 (M_W)=   K\ \cdot \ M_Q^2 (M_W)\ \cdot \ K^\dagger 
\nonumber \\
{M^{(d)}_{RR}}^2 (M_W)=   K\ \cdot \ M_D^2 (M_W)\ \cdot \ K^\dagger . 
\end{eqnarray}
{}From Eq.(\ref{doublets}) we have the universal mass for the squark doublets.
This matrix remains approximately universal at $M_W$ and hence the 
off--diagonal elements after the rotation to the SCKM basis are sufficiently
small. However the case of ${M^{(d)}_{RR}}^2$ is different. The masses
of the squark singlets, Eq.(\ref{singlets}), are not universal and hence
sizeable off--diagonal elements are generated after the rotation to the SCKM 
basis. These entries could cause problems with the bounds from mass insertions
\cite{masiero2}. However, this non--universality is diluted by the universal 
and dominant contribution from gluino to the squark mass matrices in the RGE. 
In the next sections we will analyze some CP violation observables in this
framework.

\section{EDM in models with non degenerate $A$--terms}
\label{sec:EDM}

In this section we show that, in the class of models with non
degenerate $A$--terms, the EDM of the electron and neutron can be
naturally smaller than the experimental limits,
\begin{eqnarray}
d_n &<& 6.3 \times 10^{-26}\ \mathrm{e\cdot cm}, \nonumber\\ d_e &<& 4.3
\times 10^{-27}\ \mathrm{e\cdot cm}, \label{limit}
\end{eqnarray}
even in the presence of new supersymmetric phases ${\mathcal{O}}(1)$.
As mentioned in the introduction, in the universal $A$--term scenarios we 
have severe constraints on the SUSY phases from EDMs apart from a few 
points in the parameter space where cancellations occur. In the absence of 
cancellations among different contributions $\varphi_\mu$, the phase of the 
$\mu$ term, is constrained to be ${\mathcal{O}}(10^{-2})$, while $\varphi_A$ 
is not strongly constrained \cite{CPuniv}. The cancellation 
mechanism allows for somewhat larger phases at special regions in the 
parameter space. However in this points where we could have large phases, 
we can not generate any sizable SUSY contribution to CP violation 
in the absence of new flavor structure \cite{demir,barr}. Moreover, this 
mechanism, when not justified by a symmetry argument, 
involves necessarily a certain degree of fine tuning \cite{barr}. 
In this paper, we will show that in the non--universal situation there is 
no need to restrict our parameter space to the cancellation regions to have 
large supersymmetric phases and, more important, large contributions to CP 
violation observables exist.
\vskip 0.25cm

The supersymmetric contributions to the EDM include 
gluino, chargino and 
neutralino loops. In first place, we consider the gluino
contribution which gives usually the major contribution. The
gluino contribution for the EDM of the quark $u$ and $d$ in SCKM
basis are given by
\begin{figure}
\begin{center}
\epsfxsize = 12cm
\epsffile{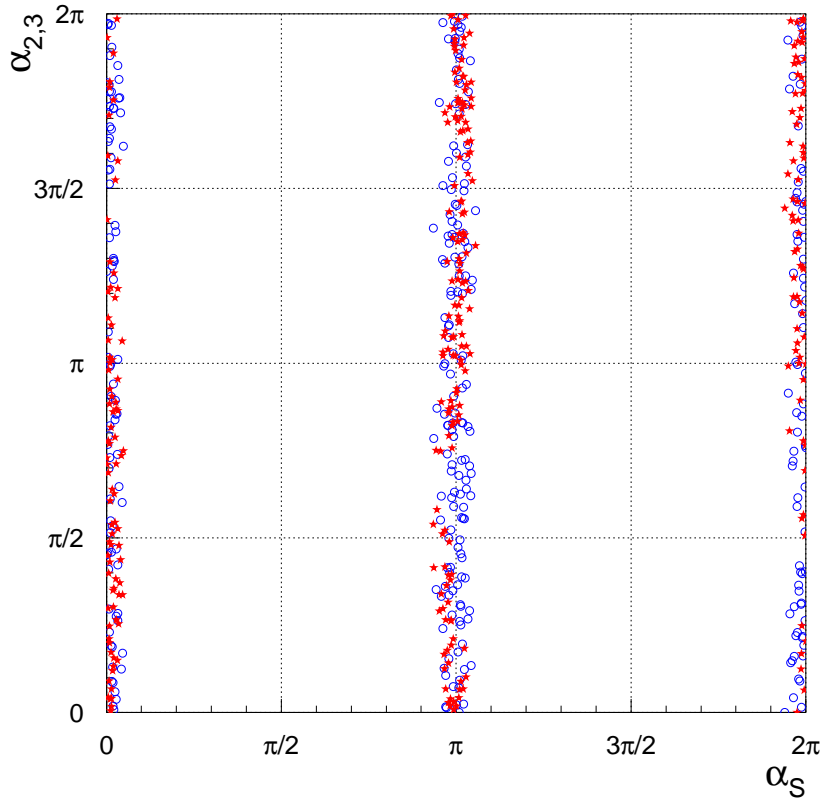}
\leavevmode
\end{center}
\caption{Allowed values for $\alpha_2$--$\alpha_S$ (open blue circles) and 
$\alpha_3$--$\alpha_S$ (red stars)}
\label{scat}
\end{figure}
\begin{eqnarray}
\label{gluinoEDM}
d^g_d/e &=& - \frac{2}{9}\ \frac{\alpha_S}{\pi}\
\frac{1}{M_{\tilde{q}}^2}\ M_1(x)\
\mathrm{Im}\{m_{\tilde{g}} (\delta_{LR}^{(d)})_{11}\}\nonumber \\ 
d^g_u/e &=&  \frac{4}{9}\ \frac{\alpha_S}{\pi}\ \frac{1}{M_{\tilde{q}}^2}\
M_1(x)\ \mathrm{Im}\{m_{\tilde{g}} (\delta_{LR}^{(u)})_{11}\}
\end{eqnarray}
where $m_{\tilde{g}}$ is complex in this model and 
$x=m_{\tilde{g}}^2/m^2_{\tilde{q}}$. The function $M_1(x)$ is given by 
$$M_1(x) = \frac{1+4 x - 5 x^2 + 4 x
\ln(x) + 2 x^2 \ln x}{2(1-x)^4}$$ and by using the
non--relativistic quark model approximation of the EDM of neutron
we can calculate it in terms of the mass insertions
$(\delta_{LR}^{(d)})_{11}$ and $(\delta_{LR}^{(u)})_{11}$.
\vskip 0.25cm

It is important to notice in Eq.(\ref{gluinoEDM}), that the relevant
phase for the gluino contribution is the relative phase between the 
gluino mass and the $L$--$R$ mass insertion \cite{kane,ibrahim}.
Thus, the physical phases entering in gluino contributions are 
$\alpha^\prime_1 = \alpha_1 - \alpha_S$, 
$\alpha^\prime_2 = \alpha_2 - \alpha_S$, 
$\alpha^\prime_3 = \alpha_3 - \alpha_S$ and 
$\varphi_{\mu}^\prime = \varphi_{\mu} -  \alpha_S$.
A very similar situation happens in the chargino contributions. These 
contributions are given by the squark and chargino mass matrices.
Hence, in the same way as with the gluino contributions, the relevant phases 
are the relative phases between chargino masses, $\alpha_1$, 
$\varphi_\mu$, and $L$--$R$ mass insertions. 
\vskip 0.25cm

As explained above, we can see from Eq.(\ref{DLR}) that these 
flavor--diagonal mass insertions tend to align with the gluino phase
(this is also true for the up squark mass matrices).
Hence, to have a small EDM it is enough to have the phases of the gauginos 
and the $\mu$ term approximately equal, $\alpha_S=\alpha_1=\varphi_\mu$. 
On the other hand, the phases of the $A$--terms are not strongly constrained
by the EDM bounds and so $\alpha_2$ and $\alpha_3$ can still be 
${\mathcal{O}}(1)$. This situation was already present even in the CMSSM
\cite{CPuniv}. Furthermore, in this case, we have the additional freedom
of independent phases for different elements of the trilinear matrix,
Eqs.(\ref{A-C1},\ref{A-C2},\ref{A-C3}).  
\vskip 0.25cm

In figure \ref{scat} we show the allowed values for $\alpha_S$, $\alpha_2$ 
and $\alpha_3$ assuming $\alpha_1=\varphi_\mu=0$. All other parameters in the 
model are scanned in the range: $60\ GeV < m_{3/2} < 300\ GeV$, 
$0.6 < \theta < 0.9$ and $0 < \Theta_i < 1$ with the constraint 
$\sum \Theta_i^2=1$. Moreover, we have imposed all the usual constraints:
\begin{itemize}
\item Squark masses above $100$ GeV with the only possible exception of the 
lightest stop and sbottom above $80 GeV$.
\item Charginos heavier than $80$ GeV
\item Branching ratio of the $b\rightarrow s \gamma$ decay, including 
supersymmetric contributions from chargino and gluino, from 
$2 \times 10^{-4}$ to $4.5 \times 10^{-4}$.
\item Gluino and chargino contributions to $d_n$ and $d_e$ independently
smaller than the phenomenological bounds.
\item Gluino and chargino contributions to $\varepsilon_K$ smaller
than $2.25 \times 10^{-3}$.
\end{itemize}
We can see that, similarly to the CMSSM situation, $\varphi_\mu$ is 
constrained to be very close to the gluino and chargino phases
(in the plot $\alpha_S \simeq 0, \pi$), but $\alpha_2$ and 
$\alpha_3$ are completely unconstrained. It is also important to notice that
we do not consider the possible cancellation regions.
In any case, these special regions would only enlarge our allowed parameter
space, mainly with larger relative phases between $\mu$ and gauginos.
However, we will see in the next section that, without these additional 
regions, large effects in CP violation observables are already present.  

\section{CP violation in the Kaon system}
\label{sec:kaon}

We have shown in the previous section, that the presence of non--universal 
$A$--terms allows the existence of large phases in the 
supersymmetry soft--breaking sector while keeping EDMs sufficiently small.
However, the important question from the phenomenological point of view is 
whether these phases are observable in other CP violation experiments 
\cite{demir}. It has been recently pointed out that, in general string
inspired SUSY models with non--universal $A$--terms, it is possible to have 
large effects in CP violation observables, and in particular in 
$\varepsilon^\prime/\varepsilon$ \cite{masiero}. In the following, we will
show that this mechanism is realized in our model and large effects are 
indeed present, while, at the same time, coping with EDM constraints. 
\vskip 0.25cm

We will mainly concentrate on the effects in the kaon system and,
in the line of Refs. \cite{abel,masiero,khalil}, we will consider the effects 
of $L$--$R$ mass insertions. In our model, as defined in section 
\ref{sec:flavor},
flavor--off-diagonality is mainly present in the down squark mass matrix.
Hence, it is clear that gluino contributions are dominant and we can 
directly apply the mass insertion bounds obtained in Ref. \cite{masiero2}. 
\vskip 0.25cm

Supersymmetric contributions to $K^0$--$\bar{K}^0$ mixing are 
mainly given by $(\delta^{(d)}_{LR})_{12}$ and 
$(\delta^{(d)}_{LR})_{21}= (\delta^{(d)}_{RL})_{12}^*$,
\begin{eqnarray}
\langle K^{0}| H_{eff} |\bar{K}^{0} \rangle_G &=& 
\frac{\alpha_s^2}{216 m_{\tilde{q}}^2} m_{K}f_{K}^{2}\ 
\Big\{ \left( (\delta^{(d)}_{LR})^2_{12} + (\delta^{(d)}_{RL})^2_{12} 
\right) \Big[44 \left(\frac{m_{K}}{m_{s}+m_{d}}\right)^{2} x\,f_6(x) 
\Big] + \nonumber \\
&&(\delta^{(d)}_{LR})_{12} (\delta^{(d)}_{RL})_{12} \Big[48 
\left(\frac{m_{K}}{m_{s}+m_{d}}\right)^{2} - 28 \Big] \,\tilde{f}_6(x) 
\Big\}
\label{dmk}
\end{eqnarray}
where $m_{K}$ and $f_{K}$ denote the mass and decay constant of the Kaon,
$x=m_{\tilde{g}}^2/m^2_{\tilde{q}}$ and the functions $f_6(x)$ and 
$\tilde{f}_6(x)$,
\begin{eqnarray}
f_6(x)=\frac{6(1+3x)\ln x +x^3-9x^2-9x+17}{6(x-1)^5}\nonumber \\
\tilde{f}_6(x)=\frac{6x(1+x)\ln x -x^3-9x^2+9x+1}{3(x-1)^5}. 
\end{eqnarray}
{}From this matrix element, the contributions to $\Delta m_K$ 
and $\varepsilon_K$
are,
\begin{eqnarray}
\label{ek}
\Delta m_{K} = 2 Re \langle K^{0} | H_{eff} | \bar{K}^{0} \rangle \nonumber \\
\varepsilon_{K}=\frac{e^{i \frac{\pi}{4}}}{\sqrt{2}}\frac{Im \langle K^0|
H_{eff} |\bar{K}^0 \rangle}{\Delta m_{K}}.
\end{eqnarray}
Similarly, these $L$--$R$ mass insertions contribute to the direct CP 
violation observable $\varepsilon^\prime/\varepsilon$. Here, the $L$--$R$
mass insertions enter mainly in the chromomagnetic penguin operators 
\cite{masiero2,buras},
\begin{equation}
Re \left(\frac{\varepsilon^\prime}{\varepsilon} \right)_G = \frac{11 \sqrt{3}}
{64 \pi}\ \frac{\omega}{ |\varepsilon|\ Re(A_0)}\ \frac{m_\pi^2 m_K^2}{f_\pi 
(m_s + m_d)}\ \frac{\alpha_s(m_{\tilde g})}{m_{\tilde g}}\   
Im\{(\delta^{(d)}_{LR})_{12}^* + (\delta^{(d)}_{LR})_{21}\}  G_0(x)
\label{epsp}
\end{equation}
with $A_i=\langle (\pi \pi)_{I=i}|H_{eff} |K^0 \rangle$, 
$\omega=Re A_2/Re A_0$. In this convention 
we have $Re A_0 = 3.326 \cdot 10^{-4}$ and  
$f_\pi= 131$ MeV. The loop function $G_0(x)$ is,
\begin{eqnarray}
G_0(x) = \frac{x(22-20x-2x^2+16x\ln(x) -x^2\ln(x)+9\ln(x))}{3(1-x)^4}~.
\label{G0}
\end{eqnarray}

Using Eqs.(\ref{ek}) and (\ref{epsp}) we can immediately calculate the 
dominant supersymmetric contributions to these observables in the presence 
of non--universal $A$ terms for a given set of initial conditions. From
\cite{masiero2}, with $x=m_{\tilde{g}}^2/m^2_{\tilde{q}}\approx 1$ the bounds 
on the $L$--$R$ mass insertions from $\Delta m_K$, $\varepsilon_K$ and 
$\varepsilon^\prime/\varepsilon$ are respectively,
\begin{eqnarray}
\label{MI-bounds}
\sqrt{|Re (\delta^{(d)}_{LR})^2_{1 2}|} < 4.4 \times 10^{-3} \cdot 
\frac{m_{\tilde{q}}(GeV)}{500} \nonumber \\
\sqrt{|Im (\delta^{(d)}_{LR})^2_{1 2}|} < 3.5 \times 10^{-4} \cdot 
\frac{m_{\tilde{q}}(GeV)}{500} \nonumber \\
\sqrt{|Im (\delta^{(d)}_{LR})^2_{1 2}|} < 2.0 \times 10^{-5} \cdot 
(\frac{m_{\tilde{q}}(GeV)}{500})^2 
\end{eqnarray}
\begin{figure}
\begin{center}
\epsfxsize = 12cm
\epsffile{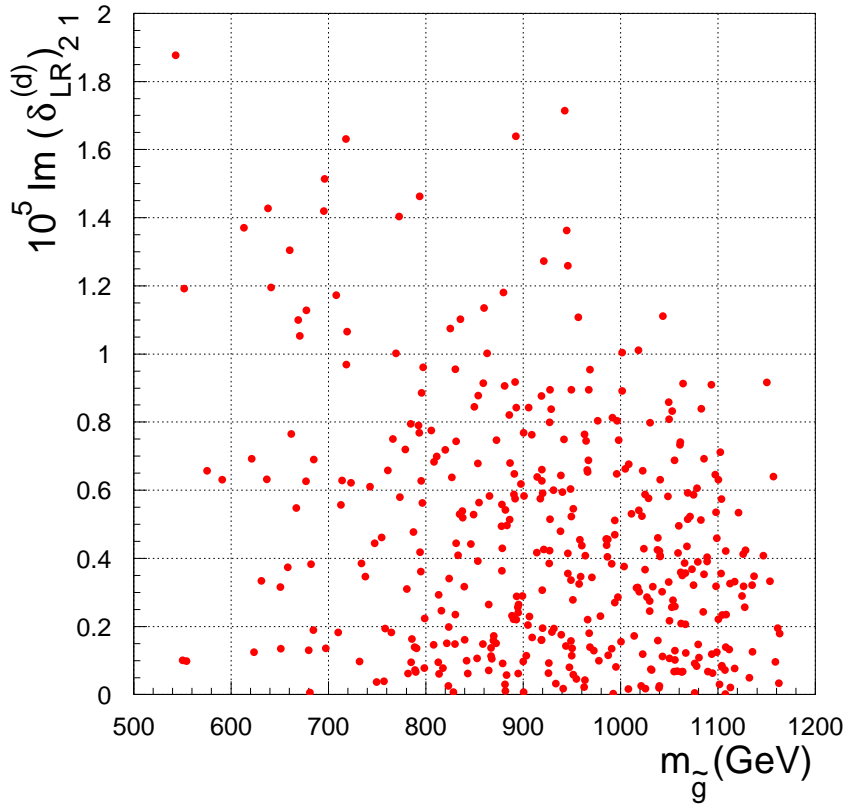}
\leavevmode
\end{center}
\caption{$(\delta_{LR}^{(d)})_{2 1}$ versus $m_{\tilde{g}}$ for experimentally 
allowed regions of the SUSY parameter space}
\label{eps'}
\end{figure}
Due to the fact that gluino amplitudes are left--right symmetric,
these bounds apply exactly the same to $(\delta^{(d)}_{RL})_{1 2}$ 
\cite{masiero2}.
This means that, for large phases, the most sensitive observable to 
non--universal $A$--terms is always $\varepsilon^\prime/\varepsilon$;
even $|{\rm Im}(\delta^{(d)}_{LR})_{2 1}^{2}| \sim 10^{-5}$ gives a 
significant 
contribution to $\varepsilon^\prime/\varepsilon$ while keeping the
contributions from this mass insertion to $\Delta m_{K}$ and $\varepsilon_K$ 
well bellow the phenomenological bounds.
In figure \ref{eps'} we show a scatter plot of values of 
$Im (\delta^{(d)}_{LR})_{2 1}$ versus the gluino mass in the same regions of 
parameter space and with the same constraints as in figure \ref{scat}.
Average scalar masses, $m_{\tilde{q}}$, are close to the gluino mass,
i.e. roughly $x \approx 1,\dots,2$.
We can see a large percentage of points are above or close to 
$1 \times 10^{-5}$, hence, sizeable supersymmetric contribution to
$\varepsilon^\prime/\varepsilon$ can be expected in the presence of 
non-universal $A$--terms. However, if we compare second and third row in 
Eq.(\ref{MI-bounds}), that is, respectively the bounds from 
$\varepsilon^\prime/\varepsilon$ and $\varepsilon_K$, it is clear
that these $L$--$R$ mass insertions can never saturate the observed value of
$\varepsilon_K$. Hence, the presence of phases in the CKM matrix is still 
required. Unfortunately, due to the large uncertainties in the theoretical
estimate of $\varepsilon^\prime/\varepsilon$, the recent experimental
measurement in KTeV and NA31 \cite{KTeV,NA31} cannot be used to constrain
this kind of model at present. In any case, the relative disagreement
between the SM predictions and the observed experimental value can be
take as a clue of new physics contributions of the kind presented 
in this paper.

\section{Conclusions}

Non--universal Supersymmetry soft breaking terms are a natural consequence
in many supergravity or string inspired SUSY models. Moreover, 
non--universality is required besides large SUSY phases to produce observable
effects in the low--energy CP violation experiments and, at the same time,
provides an efficient mechanism to allow for ${\mathcal{O}}(1)$ SUSY phases
while avoiding EDM bounds.
These features have motivated us to make a complete phenomenological analysis 
of a class of SUSY models based on type I string theory with non--universal 
scalar masses, gaugino masses and $A$--terms.
\vskip 0.25cm

Within this model, we have studied the supersymmetric contributions to 
the Electric Dipole Moments of the neutron and the electron. We find that, 
similarly to the CMSSM situation, the phase of the $\mu$ term is 
constrained to be very close to the gluino and chargino phases. However
there are still two supersymmetric phases completely unconstrained. 
This fact is completely independent of the possible existence of cancellations
between different SUSY contributions. In any case, these special cancellation
regions would only enlarge our allowed parameter space, mainly with larger 
relative phases between $\mu$ and gauginos.
\vskip 0.25cm

In the presence of these large SUSY phases, we have shown that sizeable 
supersymmetric contribution to CP observables appear. In particular, we 
have investigated the effects of these phases on the direct CP violation 
observable $\varepsilon^\prime/\varepsilon$. It has been recently suggested
that, in the presence of non--degenerate $A$--terms, large susy contributions 
to this observable are possible. Here we have demonstrated that, in this
completely defined model, this possibility is realized and a very sizeable 
fraction of the experimentally measured value can be accounted with these 
supersymmetric contributions.

\section*{ Acknowledgement}

We would like to thank D. Demir and A. Masiero for useful discussions.
O.V. thanks F.J. Botella and J. Kalkkinen for enlightening conversations
and S. K. acknowledges fruitful comments by C. Mu\~noz.  
S. K. is supported by Spanish Ministerio de Educacion y
Cultura research grant; O.V. acknowledges financial support from a Marie 
Curie EC grant (TMR-ERBFMBI CT98 3087).


\begin{thebibliography}{99}

\bibitem{2phases}
M. Dugan, B. Grinstein and L. Hall, \npb{255}{1985}{413}.
\\
S. Dimopoulos and S. Thomas, \npb{465}{1996}{23}, hep-ph/9510220.
\vskip 0.2cm

\bibitem{edm-susy}
W. Buchmuller and D. Wyler, \plb{121}{1983}{321};
\\
J. Polchinski and M. Wise,  \plb{125}{1983}{393};
\\
W. Fischler, S. Paban and S. Thomas, \plb{289}{1992}{373},
hep-ph/9205233.
\vskip 0.2cm

\bibitem{cancel}
T.~Ibrahim and P.~Nath, \plb{418}{1998}{98},
hep-ph/9707409;
\\
\prd{57}{1998}{478}; Erratum ibid \textbf{D58} (1998) 019901,
hep-ph/9708456;
\\
\prd{58}{1998}{111301}; Erratum ibid \textbf{D60} (1999) 099902,
hep-ph/9807501;
\\
M.~Brhlik, G.~Good and G.L.~Kane,\prd{59}{1999}{115004}, 
hep-ph/9810457.
\vskip 0.2cm

\bibitem{pokorski}
S.~Pokorski, J.~Rosiek and C.~Savoy, IFT preprint no. IFT-99-10A, Jun 1999,
hep-ph/9906206.
\vskip 0.2cm

\bibitem{brhlik2}
M.~Brhlik, L.~Everett, G.~Kane and J.~Lykken, 
\prl{83}{1999}{2124},
hep-ph/9905215;
\\
M.~Brhlik, L.~Everett, G.~L.~Kane and J.~Lykken, Fermilab preprint no. 
FERMILAB-PUB-99-230-T, Aug. 1999,
hep-ph/9908326.
\vskip 0.2cm

\bibitem{effective}
S.~Dimopoulos and G.F.~Giudice, \plb{357}{1995}{573},
hep-ph/9507282;
\\
A.G.~Cohen, D.B.~Kaplan and A.E.~Nelson, \plb{388}{1996}{588},
hep-ph/9607394;
\\
A.~Pomarol and D.~Tommasini, \npb{466}{1996}{3},
hep-ph/9507462.
\vskip 0.2cm  

\bibitem{abel} 
 S.~Abel and J.~Frere, \prd{55}{1997}{1623},
hep-ph/9608251.
\vskip 0.2cm

\bibitem{masiero}
A.~Masiero and H.~Murayama, \prl{83}{1999}{907},
hep-ph/9903363.
\vskip 0.2cm

\bibitem{khalil}   
S.~Khalil, T.~Kobayashi and A.~Masiero, \prd{60}{1999}{075003},
hep-ph/9903544;
\\   
S.~Khalil and T.~Kobayashi, \plb{460}{1999}{341},
hep-ph/9906374.
\vskip 0.2cm


\bibitem{non-u}
R.~Barbieri, R.~Contino and A.~Strumia, Pisa U. Report no. IFUP-TH-45-99, 
Aug. 1999, 
hep-ph/9908255;
\\
K.~Babu, B.~Dutta and R.N.~Mohapatra,
Oklahoma State University Report no. OSU-HEP-99-03, May 1999,
hep-ph/9905464.
\vskip 0.2cm

\bibitem{demir}
D.A.~Demir, A.~Masiero and O.~Vives, SISSA report no. SISSA/134/99/EP, 
Nov. 1999,
hep-ph/9911337
\\
D.~A.~Demir, A.~Masiero and O.~Vives, \prd{61}{2000}{075009},
hep-ph/9909325.
\vskip 0.2cm
                                                      
\bibitem{barr}                  
S.~Barr and S.~Khalil, \prd{61}{2000}{035005},
hep-ph/9903425.
\vskip 0.2cm

\bibitem{strings}
A.~Brignole, L.~Iba\~nez and C.~Mu\~noz, \npb{422}{1994}{125},
hep-ph/9308271;
\\
T.~Kobayashi, D.~Suematsu, K.~Yamada and Y.~Yamagishi, \plb{348}{1995}{402},
hep-ph/9408322;
\\
A.~Brignole, L.~Iba\~nez, C.~Mu\~noz and C.~Scheich, \zpc{74}{1997}{157}, 
hep-ph/9508258;
\\
For a review see e.g.\\ 
A.~Brignole, L.~Iba\~nez and C.~Mu\~noz, in Perspectives on Supersymmetry, ed.
G.~Kane, Singapore, World Scientific, 1998,
hep-ph/9707209.
\vskip 0.2cm

\bibitem{kane}
M.~Brhlik, L.~Everett, G.~L.~Kane, S.~F.~King and O.~Lebedev, Virginia Tech.
preprint no. VPI-IPPAP-99-08, Sept. 1999,
hep-ph/9909480.
\vskip 0.2cm


\bibitem{KTeV} 
KTeV Collaboration, A.~Alavi-Harati {\it et al.}, \prl{83}{1999}{22},
hep-ex/9905060.
\vskip 0.2cm

\bibitem{NA31} NA31 Collaboration (G.D.~Barr {\it et al.}\/), 
\plb{317}{1993}{233}.
\vskip 0.2cm

\bibitem{ibanez} 
L.~Iba\~nez, C.Mu\~noz and S.Rigolin, \npb{553}{1999}{43}, 
hep-ph/9812397.
\vskip 0.2cm

\bibitem{ibrahim}
T.~Ibrahim and P.~Nath, Santa Barbara U. preprint no. NSF-ITP-99-129, 
Oct. 1999, hep-ph/9910553.
\vskip 0.2cm

\bibitem{RRR}
P.~Ramond, R.~G.~Roberts and G.~G.~Ross, \npb{406}{1993}{19},
hep-ph/9303320.
\vskip 0.2cm

\bibitem{RGE}
N.~K.~Falck,
Z.\ Phys.\  {\bf C30}, 247 (1986);
\\
S.~Bertolini, F.~Borzumati, A.~Masiero and G.~Ridolfi,
Nucl.\ Phys.\ {\bf B353} (1991) 591.
\vskip 0.2cm

\bibitem{CPuniv}
T.~Falk and K.~A.~Olive, \plb{375}{1996}{196},
hep-ph/9602299;
\\
T.~Nihei, \ptp{98}{1997}{1157},
hep-ph/9707336;
\\
T.~Goto, Y.~Y.~Keum, T.~Nihei, Y.~Okada and Y.~Shimizu, \plb{460}{1999}{333},
hep-ph/9812369.
\vskip 0.2cm

\bibitem{masiero2}   
F.~Gabbiani, E.~Gabrielli, A.~Masiero and L.~Silvestrini,
\npb{477}{1996}{321},
hep-ph/9604387;
\\
J.~Hagelin, S.~Kelley and T.~Tanaka, \npb{415}{1994}{293}.
\vskip 0.2cm

\bibitem{buras}
A.~J.~Buras, G.~Colangelo, G.~Isidori, A.~Romanino and L.~Silvestrini,
\npb{566}{2000}{3},
hep-ph/9908371.

\end{thebibliography}
\end{document}